# Enhanced Pressure Dependence of Magnetic Exchange in $A^{2+}[V_2]O_4$ Spinels Approaching the Itinerant Electron Limit.

S. Blanco-Canosa,[1] F. Rivadulla,[1]* V. Pardo,[2,3] D. Baldomir,[2,3] J.-S. Zhou,[4] M. García-Hernández,[5] M. A. López-Quintela,[1] J. Rivas,[2] J. B. Goodenough.[4]

*Departamento de Química-Física[1] y Física Aplicada[2], Universidad de Santiago de Compostela, 15782-Santiago de Compostela, Spain.*
[3]*Instituto de Investigaciones Tecnológicas, Universidad de Santiago de Compostela, 15782-Santiago de Compostela, Spain*
[4]*Texas Materials Institute, The University of Texas at Austin, USA.*
[5]*ICMM, Madrid, Spain.*

**We report a systematic enhancement of the pressure dependence of $T_N$ in $A^{2+}[V_2]O_4$ spinels as the V-V separation approaches the critical separation for a transition to itinerant-electron behavior. An intermediate phase between localized and itinerant electron behavior is identified in $Zn[V_2]O_4$ and $Mg[V_2]O_4$ exhibiting mobile holes as large polarons. In $Zn[V_2]O_4$, cooperative ordering of V-V pairs below a $T_s \approx T_N$ does not totally suppress the $V^{3+}$-ion spins at ambient pressure, but makes $T_N$ to decrease with pressure. Our results demonstrate that $Zn[V_2]O_4$ and $Mg[V_2]O_4$ are less localized than previously thought.**

The observation in 1957 by Bongers[1] of a sharp drop in the paramagnetic susceptibility of layered $LiVO_2$ on cooling through $T_t = 490$ K led to the prediction[2] that it was caused by the formation of $V^{3+}$-ion triangular clusters (trimers) within the $VO_2$ planes since the two 3d electrons of a $V^{3+}$ ion would allow formation of two V-V homopolar bonds per $V^{3+}$ ion across shared octahedral-site edges oriented at 60º with respect to one another. At that time, it was not yet appreciated that the transition from localized to itinerant electrons is first-order, which allows cation-cluster formation to occur on approaching crossover from either the localized-electron or itinerant-electron side. However, it was understood that a transition to cation clustering would occur only at this crossover, and therefore a critical V-V separation $R_c \approx 2.94$ Å for the transition was estimated for an array of octahedral-site $V^{3+}$ ions sharing octahedral-site edges with 6 $V^{3+}$-ion nearest neighbors[2]. The single-valent spinels $A[V_2]O_4$ (A = divalent cation) also contain V-V interactions between 6 nearest-neighbor $V^{3+}$ ions sharing octahedral-site edges, but in a 3-D rather than a 2-D array. Since the $A[V_2]O_4$ spinel family also approaches $R_c$ from the localized-electron side, we decided to reinvestigate these spinels to see whether precursors to the transition to itinerant-electron behavior could be identified for V-V bonding as has been done in $LaMnO_3$ for Mn-O-Mn bonding.[3]

The approach to this itinerant-electron limit from the localized side, could be monitored through the volume dependence of the Néel temperature, i.e. the Bloch parameter

$$\alpha \equiv - d(lnT_N)/d(lnV) \qquad (1)$$

which can be determined by measuring $dlnT_N/dP$ and the compressibility $d(lnV)/dP$. Bloch[4] observed that a large number of localized-electron, single-valent antiferromagnetic oxides have an $\alpha \approx 3.3$. This value is so general that even small deviations to higher values can be used to signal the approach to itinerant-electron behavior from the localized-electron side.[3] The value of $\alpha \approx 3.3$ has been justified theoretically provided the Coulomb energy U entering the exchange coupling

$$J \sim t^2/U \qquad (2)$$

is independent of pressure. An $\alpha \approx 3.3$ follows since the spin-dependent expectation value t for a charge transfer between sites is proportional to the overlap integral for the donor and acceptor orbitals on neighboring atoms, which varies sensitively with the interionic distance.

In $A^{2+}V_2O_4$ (A= Cd, Mn, Co, Zn, Mg), the V-V distance can be varied in a systematic fashion by changing the $A^{2+}$ cation at the tetrahedral site. Therefore, this series provides an exceptional system to study the breakdown of perturbation formula of superexchange theory in a single-valent system with direct metal-metal bonding as the itinerant electron limit is approached. This system reduces the inherent complexity of similar studies in perovskites, where superexchange introduces a dependency on the charge-transfer gap, $\Delta$, which has a variation with volume that is uncertain.

Single-phase, polycrystalline $A[V_2]O_4$ spinels were synthesized by solid-state reaction in evacuated quartz ampoules. Structural parameters were derived from X-ray diffraction patterns by Rietveld analysis. Magnetic measurements under pressure up to P = 10 kbar were obtained by using a Be-Cu cell. Lattice compressibility was measured by fitting the pressure dependence of the lattice volume (up to 50 kbar) with the Birch-Murnaghan equation. *Ab initio* density-functional-theory (DFT) calculations were performed within a full-potential, all-electron approximation with the WIEN2K software.[5] Strong correlations were modeled by means of a hybrid DFT scheme for correlated electrons, the PBE0 functional.[6]

The octahedral-site ions of a spinel form a pyrochlore lattice, which is geometrically frustrated for antiferromagnetic interactions. Consequently $T_N$ is reduced considerably from the value expected on the basis of the Weiss constant $\theta_w$ of the Curie-Weiss paramagnetic susceptibility, and an enormous $\theta_w/T_N$ ratio results.[7] However, the geometric frustration remains volume-independent provided the lattice symmetry is not changed by the pressure. Therefore, Bloch's rule, $\alpha \approx 3.3$, should remain applicable to spinels where octahedral-site interactions are dominant. In order to test this conclusion, we can measure $\alpha$ for the $A[Cr_2]O_4$ spinels since they have an $R > R_c$ with an estimated $R_c \approx 2.84$ Å. In order, to maximize $R > R_c$, we have chosen to measure $\alpha$ for $Mn[Cr_2]O_4$ having an R = 2.93(1). In this spinel, given the electronic configuration of $Cr^{3+}$, the Cr-Cr interactions should dominate over the Mn-O-Cr interactions[8]. By fitting total energy calculations to a Heisenberg model, we calculated a $J_{Cr-Cr}/J_{Cr-Mn} \approx 3$, confirming our initial hypothesis.

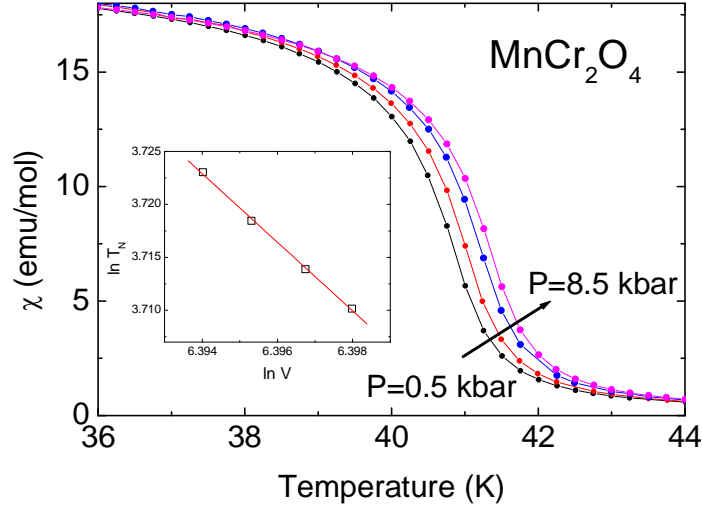

Figure 1: Pressure and temperature dependence of the magnetization (H=100 Oe) around $T_N$ in Mn[$Cr_2$]$O_4$. Inset: Fit of the experimental data to equation (3).

Magnetization curves of Mn[$Cr_2$]$O_4$ measured around $T_C$ for different pressures are shown in Fig. 1. The finite magnetization below $T_C$ is due to the formation of a complex spiral-spin configuration in this frustrated ferrimagnet.[9] Taking $T_C = T_N$ for the Cr-Cr interactions, the data of Fig. 1 give a $T_N^{-1} dT_N/dP = 1.7 \times 10^{-3}$ kbar$^{-1}$; with a measured compressibility $\kappa = 4.95 \times 10^{-4}$ kbar$^{-1}$, an $\alpha = 3.4(1)$ is obtained for Mn[$Cr_2$]$O_4$. This value follows nicely the Bloch rule for the Cr-Cr interactions, which validates our use of $\alpha$ as an indicator whether any of the A[$V_2$]$O_4$ spinels approach the transition to itinerant electron behavior from the localized-electron side.

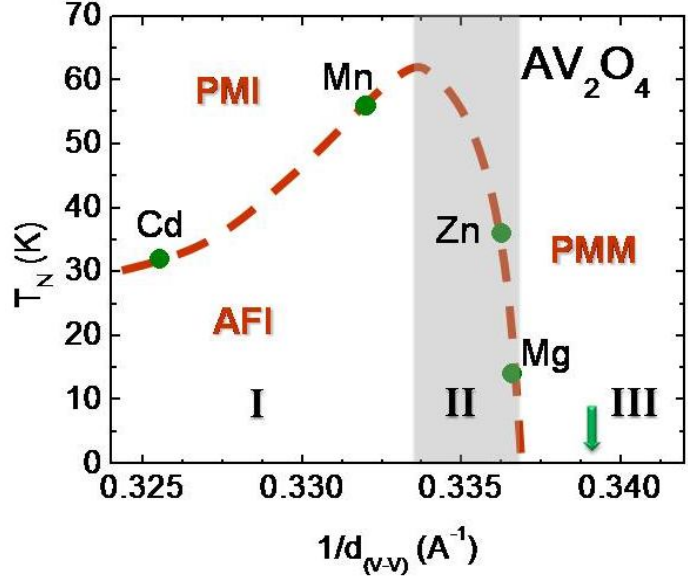

Figure 2: Magnetic phase diagram of $A[V_2]O_4$ spinels as a function of the V-V distance. The dotted line is a guide to the eye. The vertical arrow marks the estimated critical distance for itinerant-electron behavior (see the text).

Figure 2 shows the evolution of $T_N$ with the inverse V-V separation, $1/d_{V-V}$, for the $A[V_2]O_4$ spinels. Such an evolution has been predicted for the approach to crossover from the localized-electron side.[10] As the overlap integral within t of equation (2) increases with decreasing temperature, the energy U must decrease or remain constant. Therefore, within the localized-electron superexchange perturbation theory represented by equation (2), $T_N$ should increase with decreasing V-V separation, *i.e.* increasing $1/d_{V-V}$. Therefore, we assign $Cd[V_2]O_4$ and $Mn[V_2]O_4$ to a localized-electron regime, Phase I. The spinels $Zn[V_2]O_4$ and $Mg[V_2]O_4$, which fall in the shaded region of Fig. 2, we assign to a Phase II intermediate between Phase I and an itinerant-electron Phase III. Rogers *et al.*[11] investigated the possibility of reaching phase III in the $A[V_2]O_4$ spinels. Although this early attempt failed, they found that the activation energy for the resistivity decreased progressively as the V-V separation was reduced.

Although A-site $Mn^{2+}$ has a magnetic moment, the inclusion of $Mn[V_2]O_4$ in Fig. 2 is fully justified. The $t^2e^0$ configuration of a $V^{3+}$ ion allows for a V-V interaction across the shared octahedral-site edges that is strong relative to the Mn-O-V interactions[12] as in Mn[Cr$_2$]O$_4$. Our *ab-initio* calculations fit to a Heisenberg model, as we did for Mn[Cr$_2$]O$_4$, yielded J$_{V-V}$ =20 K, J$_{Mn-V}$ = 9 K, and J$_{Mn-Mn}$ = 0.3 K, this latter value agreeing with that found experimentally for Mn[Al$_2$]O$_4$.[13] Comparison of the T$_C$ for Mn[Cr$_2$]O$_4$ ($\approx$ 50 K) and Mn[Fe$_2$]O$_4$ ($\approx$ 570 K) shows that half-filled e orbitals on the octahedral-site Fe$^{3+}$:t$^3$e$^2$ ions enhance greatly the Mn-O-Fe interactions relative to weakened Fe-Fe interactions.

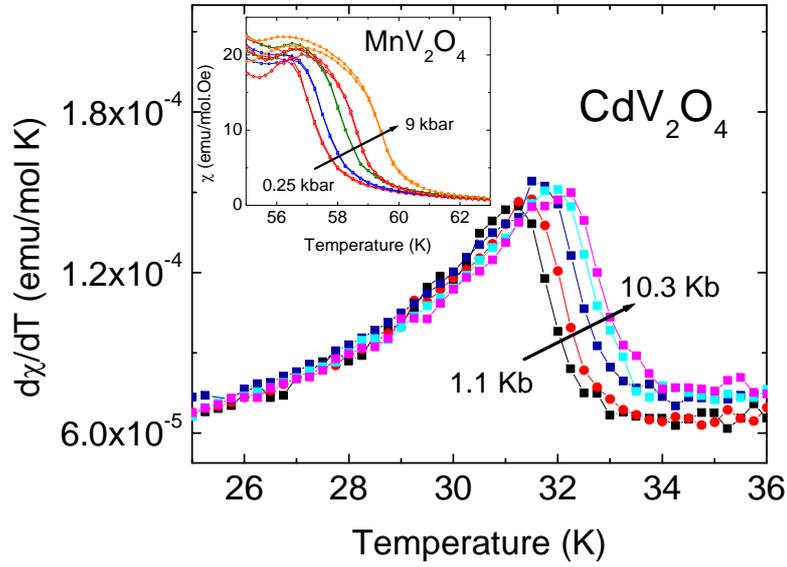

Figure 3:. Pressure dependence of T$_N$ in Cd[V$_2$]O$_4$. We represent the derivative of $\chi$(T), in order to determine T$_N$(P) accurately. The inset shows $\chi$(T) for MnV$_2$O$_4$ at different pressures around T$_N$.

Given the applicability of Bloch's rule to spinels with a dominant superexchange interaction between octahedral-site ions, we will now demonstrate the progressive increase of $\alpha$ from the Bloch value of 3.3 as the V-V separation decreases in the A[V$_2$]O$_4$ spinels. Figure 3 shows the pressure dependence of T$_N$ for Cd[V$_2$]O$_4$ and Mn[V$_2$]O$_4$; the data give a T$_N^{-1}$dT$_N$/dP = 2.7(4)×10$^{-3}$ kbar$^{-1}$ for Cd[V$_2$]O$_4$ and an anomalously large 5.6(3)×10$^{-3}$ kbar$^{-1}$ for Mn[V$_2$]O$_4$. We have measured a compressibility $\kappa$ = 5.88×10$^{-4}$ kbar$^{-1}$ for Mn[V$_2$]O$_4$, a little larger that the $\kappa$ = 4.95×10$^{-4}$ kbar$^{-1}$ obtained for Mn[Cr$_2$]O$_4$. Given a compressibility $\kappa$ =6.6×10$^{-4}$ kbar$^{-1}$ for Cd[V$_2$]O$_4$,[14] we obtain the Bloch parameters $\alpha \approx 4.1(1)$ for Cd[V$_2$]O$_4$ and $\alpha \approx 9.9(1)$ for Mn[V$_2$]O$_4$. The value of $T_N^{-1}dT_N/dP$ for Mn[V$_2$]O$_4$ was obtained with two different measurements, giving the same $\alpha$, the largest reported so far in an oxide. Whether this large value of $\alpha$ reflects an anomalous compressibility near $T_N$ because of a large exchange striction has yet to be examined. Nevertheless, these results signal that the assumption of a pressure-independent energy U in the localized-electron superexchange theory behind equation (2) is breaking down at the transition from Phase I to Phase II of Fig. 2.

An important consequence of this study is that in the transitional Phase II, neither the localized-electron nor the itinerant-electron model is applicable; the $V^{3+}$-3d electrons are in a state that, at low temperatures, approaches a quantum phase transition (QPT). If the transition from localized to itinerant electronic behavior is first-order, lattice instabilities can be anticipated. These instabilities would dress mobile holes in a multi-cation cluster, *i.e.* give a large rather than a small polaron as occurs in the parent La$_2$CuO$_4$ phase of the copper-oxide La$_{2-x}$Sr$_x$CuO$_4$ superconductors.[15] They would also induce a phase transition by a cooperative formation of V-V pairs bonded by homopolar bonds as in LiVO$_2$. Large polarons are distinguishable from small polarons in a measurement of the thermoelectric power.

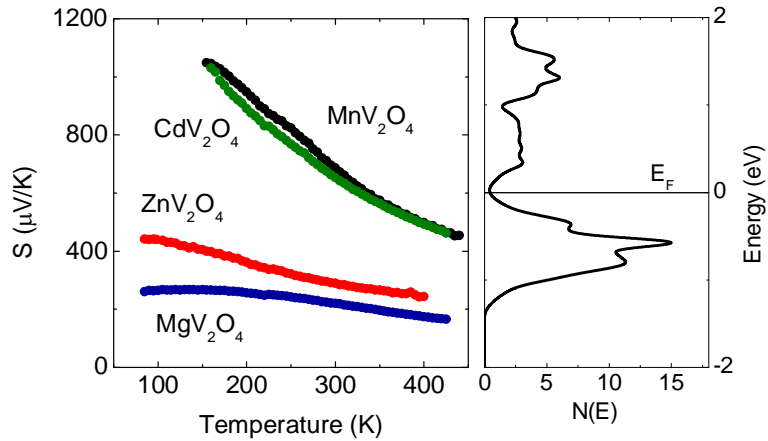

Figure 4: Thermoelectric power of the series A[V$_2$]O$_4$. Reduction in the absolute value and loss of typical activated behaviour is evident in Zn and Mg samples, both in the shadow area of the phase diagram in Fig. 2. On the right panel, calculated total density of states is shown for $MgV_2O_4$ under an applied pressure P= 8 GPa. At such a high pressure, the V-V distance is predicted to be 2.931 Å and the system to be metallic.

The temperature dependence of the thermoelectric power, S(T), is shown in Fig. 4 for the $A[V_2]O_4$ spinels. The large value of S(T) shows that there are only low density mobile holes in these stoichiometric oxides. Thermally activated behavior, typical of trapping of small polarons at the defects that created them, is observed for $Cd[V_2]O_4$ and $Mn[V_2]O_4$. $Zn[V_2]O_4$ and $Mg[V_2]O_4$, on the other hand, exhibit a strongly reduced activation energy that, in the case of $Mg[V_2]O_4$, shows a non-activated hole conduction at lowest temperatures that is reminiscent of the behavior of the minority-spin charge carriers of $Fe_3O_4$ between the Verwey transition ($\approx$ 120 K) and room temperature.[16] $Zn[V_2]O_4$ and $Mg[V_2]O_4$ appear to contain untrapped mobile holes that move with a hopping time $\tau_h \approx \omega_R^{-1}$ at the periphery of a large-polaron cluster; $\omega_R$ is the frequency of the optical-mode lattice vibrations.

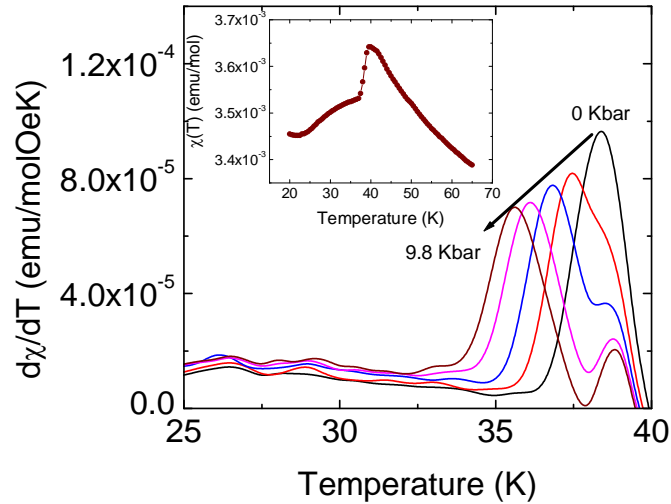

Figure 5: Pressure dependence of $\chi(T)$ for $Zn[V_2]O_4$. We present the derivative of $\chi(T)$ for the sake of clarity, as the two-peak structure is clearer than in the direct representation. Inset: ambient pressure $\chi(T)$ curve for $Zn[V_2]O_4$ measured at H=1 T. The drop at ~40 K marks $T_S=T_{OO}$ (see text).

Consistent with this indication that $Zn[V_2]O_4$ and $Mg[V_2]O_4$ belong to a transitional phase is the observation in both spinels of a first-order tetragonal (c/a < 1) to cubic transition at a temperature $T_s$. These transitions are characteristic of a cooperative ordering of strong V-V bonding at the expense of empty V-V bonds as the crossover to itinerant-electron behavior is approached from the localized-electron side. Although the driving force for these phase transitions is still under discussion,[17,18,19,20] our results show that models starting with a U >> t are not appropriate. With an angle α = 60° between nearest V-V bonds for the yz and zx orbitals, the tetragonal distortion may be characterized by a Peierls dimer formation in 1-D chains along [011] and [101] directions as pointed out by Khomskii and Mizokawa.[20] The situation is completely analogous to the formation of $V_3$ trimers within the $VO_2$ planes of $LiVO_2$. Cooperative ordering of metal-metal bonds is a common phenomenon at the crossover from localized to itinerant electronic behavior as is illustrated by the spinels $Cu[Ir_2]S_4$ and $Mg[Ti_2]O_4$[21,22] as well as by numerous molybdates.[23]

Experimental NMR, MSR, and neutron-diffraction studies[24,25] have demonstrated that the V-V pairing in the tetragonal phase does not suppress totally the spin on a vanadium ion nor remove geometric frustration completely, leading to a short-range magnetic order below $T_N \approx T_S$ and an incommensurate spin-density wave. This finding may be characteristic of homopolar bonding where the metal-metal separation remains R ≥ $R_C$. Specific-heat measurements (not shown) reveal a second anomaly at *ca*. 30 K that can be associated with long-range magnetic order, following ref. [26]. Figure 5 shows the effect of pressure on the magnetic susceptibility of $Zn[V_2]O_4$. At ambient pressure, the susceptibility drops sharply on cooling through $T_s$ = 38 K due to the spin pairing in the V-V bonds. Under pressure, the two-peak structure of the plot of dχ/dT vs T associated with $T_s$ and $T_N$ is resolved. Given c/a<1 in the tetragonal phase, the peak that decreases with pressure is ascribed to $T_N$. If the residual spins on the V-V pairs are reduced by pressure, the long-range magnetic ordering temperature $T_N$ is reduced. Long-range magnetic order and any $V^{3+}$-ion spin may be completely suppressed below $T_s$ in $Mg[V_2]O_4$.

Finally, given the experimental compressibility of 5.88 × $10^{-7}$ $bar^{-1}$ and an $R_C$ ≈ 2.94 Å (marked by an arrow in Fig. 2), we predict a QPT at 7 GPa. Our *ab initio* calculations predict an overlap of the valence and conduction bands, and hence metallic conductivity at *c.a.* ~ 6.5 GPa. Experiments at higher pressure would reach phase III in a single valence system with direct metal-metal bonding.

In summary, we have determined the pressure dependence of $T_N$ for several $A^{2+}[V_2]O_4$ spinels to show that as the V-V separation is reduced, there is a breakdown of the localized-electron model for the 3d electrons. The breakdown is first marked by a pressure dependence of the Hubbard U in $Mn[V_2]O_4$ and then by long-range ordering of V-V pairs in $Zn[V_2]O_4$ and $Mg[V_2]O_4$. However, the $V^{3+}$-ion spins of the V-V pairs are not totally suppressed by the spin pairing within the homopolar bonds so long as an R > $R_c$ is retained. However, the introduction of V-V pairs below $T_s$ reduces the long-range ordering temperature $T_N$ and makes $T_N$ fall sharply with increasing pressure. The experimental data indicate partial electronic delocalization in the V-V pairs of $Zn[V_2]O_4$ supporting the orbitally driven Peierls state proposed by Khomskii and Mizokawa.[20] Finally, a*b initio* calculations predict a gap-closing and a crossover to metallic state in $Mg[V_2]O_4$ and $Zn[V_2]O_4$ would occur at a pressure of ~ 6.5 GPa.

**Acknowledgments.** Prof. D. Khomskii and Prof. M. M. Abd-Elmeguid, are acknowledged for discussion and critical reading of the manuscript. Financial support from Xunta de Galicia (Project: PXIB20919PR), MEC (Project: NMAT2006-10027) and Ramón y Cajal (F.R.), and FPU (S.B.-C.) programs, are also acknowledged. We also thank the CESGA (Centro de Supercomputación de Galicia) for computing facilities.